\documentclass{aa} 
\usepackage{graphicx}
\usepackage{natbib}
\bibpunct{(}{)}{;}{a}{}{,}
\usepackage{txfonts}
\begin{document}
   \title{Detection of an optical filament in the Monogem Ring
   \thanks{Based on observations collected at the 1.82 m telescope of the
   Padua Astronomical Observatory at Asiago and
   at the 2 m Schmidt telescope of the
   Th\"uringer Landessternwarte Tautenburg.}}

   \subtitle{}

   \author{Ronald Weinberger\inst{1},
           Sonia Temporin\inst{1},
       \and
       Bringfried Stecklum\inst{2}
          }

   \offprints{ronald.weinberger@uibk.ac.at}

   \institute{Institut f\"ur Astrophysik, Universit\"at Innsbruck,
              Technikerstra{\ss}e 25, 6020 Innsbruck, Austria\\
              \email{ronald.weinberger@uibk.ac.at}\\
          \email{giovanna.temporin@uibk.ac.at}
         \and
              Th\"uringer Landessternwarte Tautenburg,
          Sternwarte 5, 07778 Tautenburg, Germany\\
             \email{stecklum@tls-tautenburg.de}
             }

   \date{Received September ??, 2005; accepted ?? ??, ????}

\authorrunning{R. Weinberger et al.}
\titlerunning{A new optical filament in the Monogem Ring}

   \abstract{
   The Monogem Ring is a huge  bright soft X-ray enhancement with a
   diameter of $\sim$ 25$\degr$. 
   This 0.3 kpc distant structure 
   is a peculiar Galactic supernova remnant in that it is obviously 
   visible only in X-rays, due to its expansion into a region of extremely low
   ambient density: hence, practically no optical emission or a neutral
   \ion{H}{i} shell
   was expected to be detectable. -  Here we report on the discovery of a
   very
   faint arc-like nebula on a POSS\,II R film copy,
   at the south-eastern borders of the MR.
   Spectroscopy revealed this filament to have a
   very large [\ion{S}{ii}]$\lambda$
6716+6731/H$\alpha$ ratio of up to $\sim$ 1.8, indicating
   shock excitation,  and a low density of N$_{\rm e}$ $<$100
   cm$^{-3}$. There is no hint of [\ion{O}{iii}] emission in the spectra.
   On deep wide-field direct images in H$\alpha$ and in
   [\ion{S}{ii}] the nebula appears as a $\sim$ 20\arcmin \,long, thin
   ($\sim$ 1\arcmin), structured filament, stretching N-S. We
   believe that this filament belongs to the MR and
   became visible due to the interaction of the expanding remnant with a mild density
   increase in the
interstellar medium. Only one other possible optical filament of
the MR has been reported in the literature, but no spectrum was
provided.

   }

   \maketitle
%

\section{Introduction}

The \object{Monogem Ring} (MR) is a $\sim$ 25\degr -diameter supernova remnant
 (SNR) centered at ($\ell$, $b$) $\approx$ (203\degr, +12\degr),
 visible as a bright, diffuse, soft X-ray enhancement in the
 Monoceros and Gemini constellations. Recently, this $\sim$ 300 pc
 distant structure has attracted considerable attention, as a
 very important source of high-energy cosmic
 rays and a possible responsible for  the sharp knee in the
 cosmic ray energy spectrum at $\sim$ 3 PeV \citep{ew,cmg,tbb,kky,kz}.

The MR is 
interesting in another respect too. According to a
detailed modeling of \citet{psa}, this SNR is in the
adiabatic stage of evolution: At a distance of 300 pc and with an
average temperature of log($T$/K) = 6.15, among other things the
initial ambient density was found to be extremely low,
5.2\,10$^{-3}$ cm$^{-3}$ only. The thermal pressure of the
X-ray-emitting gas was calculated by these authors to be
4.5\,10$^4$ K cm$^{-3}$, only 2-4 times higher than the estimated
pressure of the undisturbed interstellar medium (ISM). Hence, as
this SNR continues to evolve, it probably will reach pressure
equilibrium with the ISM before it reaches the radiative phase of
SNR evolution. As a consequence of this very low ambient density,
optical filaments (and a neutral shell) should be particularly
difficult to detect in the MR, as \citet{psa}
emphasize.

Interestingly, in a so far unnoticed paper published more than 2
decades ago, \citet{rw} reported on a previously
unknown faint interstellar filament found on a deep U-plate.
Although no spectrum was taken, these authors provided arguments
that this about 1\degr \,long thin filament, with a maximum near 
R.A. 06:30.5, Dec. +09.75 (B1950.0), represents the outcome
of [\ion{O}{ii}] line emission due to a slow shock  and appears to
be an optically visible part of the south-western region of the
Monogem Ring. There is another (passing) mention of optical
emission from the MR in the literature, referring to faint
extended optical emission in the Wisconsin H-Alpha Mapper (WHAM)
survey \citep{hrt}; due to its angular resolution of
1\degr \, any sharp filaments could of course not be detected.

Here we report on a new, very faint arc-like nebula visible on an R
film copy of the Second Palomar Observatory Sky Survey (POSS\,II).
It was detected in the course of our search for galaxies in the
zone-of-avoidance \citep{ssw}. On deep wide-field
images taken with the 2m Tautenburg Schmidt telescope it appears
as a long, structured, thin filament. Spectra obtained with the
1.8m telescope on Cima Ekar in Asiago show this filament to be
excited by a slow shock. Its location strongly suggests that we
deal with optical emission from the Monogem Ring.


\section{Observations}
Broad band I (2$\times$ 180 sec) and narrow band H$\alpha$
(2$\times$ 1200 sec) and [\ion{S}{ii}] $\lambda$ 6716,6731 \AA\
(2$\times$ 1200 sec) exposures were acquired in March 2005,  at
the Th\"uringer Landessternwarte Tautenburg (Germany),
with the 2 m telescope in the Schmidt mode, for which the 
correction plate limits the aperture to 1.34 m.
The bandpass width (FWHM) of the H$\alpha$ and [\ion{S}{ii}] 
filters is 9.7 nm.
The pixel size of the 2k$\times$2k SITe CCD chip, 24$\times$24 $\mu$m, 
and the plate scale of 51.4 arcsec mm$^{-1}$ give a field of view of
42\arcmin\ $\times$ 42\arcmin\ with an image scale of 1.23 arcsec pixel$^{-1}$. 
While no structure was detected in the
I-band image, a curved filament of $\sim$  20\arcmin\ extent in
North-South direction was detected in the two narrow-band filters.

Optical spectra across the brighter part of the filament,
centered at R.A. 07:28:38, Dec. 05:09:55 (J2000), were
obtained during three observing runs in November and December 2004
and in January 2005, at the 1.82 m telescope of the Asiago
Observatory (Italy) with the Asiago Faint Object Camera and
Spectrograph (AFOSC) combined with a TEK 1024 thinned CCD chip, 
giving a spatial scale of 0.47 arcsec
pixel$^{-1}$. Low resolution long-slit spectra in the wavelength
range 4000 - 7800 \AA\  (dispersion 4.22 \AA\ pixel$^{-1}$,
spectral resolution $\sim$  29 \AA) were obtained in the first two
runs under unfavourable ($\sim$  3\arcsec) seeing conditions with
total integration times of 3600 and 2900 sec. A higher resolution
spectrum in the wavelength range 6250 - 8050 \AA\ (dispersion 1.77
\AA\ pixel$^{-1}$, spectral resolution $\sim$  5.4 \AA) was obtained
in the third run under spectrophotometric conditions and $\la$
2\arcsec\ seeing with an exposure time of 3600 sec. A slit width
of 1\farcs69 was chosen. The emission lines H$\alpha$,
[\ion{N}{ii}] $\lambda\lambda$ 6548,6583 \AA, and [\ion{S}{ii}]
$\lambda\lambda$ 6716,6731 \AA\ were clearly detected in all
observations over an extent of $\sim$  55\arcsec\ along the slit,
without any visible underlying continuum emission.
The [\ion{O}{i}] $\lambda$ 6300 emission line is also visible,
but is very weak and suffers from contamination by bright night-sky
lines.

The data reduction proceeded in the usual way. Standard
IRAF\footnote{IRAF is distributed by the National Optical
Astronomy Observatories, which are operated by the Association of
Universities for Research in Astronomy, Inc., under cooperative
agreement with the National Science Foundation.} packages were
used for bias subtraction, flat fielding, wavelength
linearization, flux calibration and subtraction of the
sky-background, while the package {\sc L.A.COSMIC} \citep{vd}  
was used for cosmic ray cleaning. Flux calibration was
performed by use of spectrophotometric standard stars observed
during the same nights as the target object.

Emission lines were measured by Gaussian fitting. No correction
for foreground extinction was applied, however the emission lines
under consideration are quite close in wavelength, therefore their
ratios are not expected to be significantly affected by extinction
(see Sect. 3.2).

\section{Results and discussion}
\subsection{Morphology}
A slightly curved filament of $\sim$ 20\arcmin\ extent in
North-South direction was detected in direct images, with the two
narrow-band filters. It runs from R.A. 07:28:28.9, Dec. +05:13:04
to 07:28:19.7, +04:57:13 (J2000) with a brightness maximum near
R.A. 07:28:38, Dec. +05:10. The structure can be roughly
approximated with an arc of circumference of radius
$\sim$ 7\arcmin, except for the southernmost $\sim$ 6\arcmin \, part
of the filament, which seems to
 be arranged on a straight line. On the northern tip, the filament is
crossed by a small arclet $\sim$ 5\farcm5 long and $\la$1\arcmin
\,wide. Upon closer examination one gains the impression that
close in projection to the (unrelated) bright star located
1-2\arcmin \, South-West of the brightness maximum (the F4V star
\object{GAT 1096}; B = 9.92, V = 9.51), the layers of the filament twist:
e.g. the westernmost layer of the southern part of the filament
becomes the easternmost layer in the northern part of the
filament. A composite image obtained from the exposures in the
three bands is shown in Fig.~\ref{truecolor}.

   \begin{figure*}
   \centering
Figure available as jpeg
\vskip 1cm   
\caption{Composite I-band, H$\alpha$, [\ion{S}{ii}] image of the filament, shown in
   logarithmic scale with a lower cutoff of 0.5$\sigma$, colour coded
   so that I appears red, H$\alpha$ emission appears blue, and [\ion{S}{ii}] emission appears
   green. The tweezers-like E-W-structure in the lower half of the image is not real.
   The field of view is $\sim$  16\farcm8 $\times$ 24\farcm8, centered at
   R.A. 07:28:27.3, Dec. $+$05:06:43. North is on top, East to the left.}
              \label{truecolor}
    \end{figure*}

Using {\it{SkyView}}\footnote{http://skyview.gsfc.nasa.gov} we found  
no counterparts of this filament at other wavelengths. On IRAS maps, 
for example, there is no extended
emission visible near or along this structure. In the optical, on
POSS\,II R and B film copies, no sign of dust extinction is
traceable. An examination of the Southern H-Alpha Sky Survey Atlas
\citep[SHASSA,][]{gau}, which covers the southern hemisphere sky up to
declinations of +15\degr \, shows no emission worth mentioning at
or within a few degrees within the location of the filament.
Hence, the filament and the northern arclet seem to be isolated
structures, at least  up to the available brightness limits.

\subsection{Spectroscopy}
The filament clearly is an emission nebula. Emission line ratios
derived from the low resolution spectra are in agreement within
errors with those from the high resolution one, therefore we
present here only the latter. The slit position is marked on the
[\ion{S}{ii}] image in Fig.~\ref{slit}. Its position is centred at
R.A. 07:28:38, Dec. 05:09:55 (J2000),
 i.e. at Galactic coordinates $\ell$ = 212$\fdg$5390,
 $b$ = +10$\fdg$6193.

   \begin{figure*}
   \centering
\vskip 1cm
Figure available as jpeg   
\caption{A detail of the [\ion{S}{ii}] image, centred at R.A. 07:28:35.0,
   Dec. $+$05:09:12 (J2000), showing the position of the slit
   during spectroscopic observations. The presence of two filaments is clearly visible.}
              \label{slit}
    \end{figure*}

 A tentative
detection of the H$\beta$ emission line at 1.4$\sigma$ level in
the low resolution spectrum can be used as an upper limit to the
line flux. This gives a lower limit to the H$\alpha$/H$\beta$
ratio of $\sim$ 5, thus implying a lower limit to the foreground
extinction E($B-V$) $\sim$ 0.5 mag for an assumed intrinsic ratio
H$\alpha$/H$\beta$ = 3.0.
This value of foreground extinction is significantly higher than
that implied by dust maps derived from IRAS and DIRBE far infrared
data in this region of the sky. In fact, using the dust maps
of \citet{sch} with a scale of 2.37 arcmin$^2$ pixel$^{-1}$ 
and interpolating from the four nearest pixels around the 
position $\ell$ = 212.5, $b$ = 10.6, we obtained E($B-V$) = 0.07 mag.
Since the spatial extent of H$\alpha$ emission in our
spectra (along the slit) is $\lesssim$ 1\arcmin\, a possible explanation
to this discrepancy might be the presence of a local
patch of dust associated to the filament, small and/or cold enough
not to be visible in the dust maps. Another possible explanation
dealing with the physical conditions of the emitting gas is proposed
later in this Section.

\begin{figure}
   \centering
   \includegraphics[width=\linewidth]{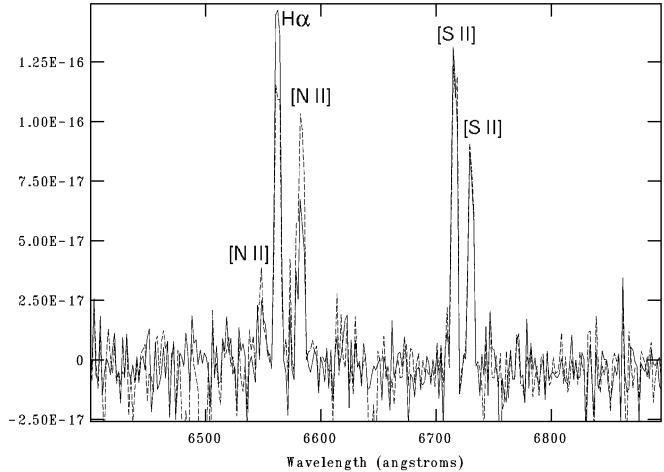} 
   \caption{Spectra of the western and eastern portions of the filament
   (solid and dashed line, respectively), showing the different relative
   intensities of the emission lines. Flux units are ergs cm$^{-2}$ s$^{-1}$ \AA$^{-1}$.}
              \label{spectra}
    \end{figure}

A gradient in the emission-line ratios of the filament is present
along the slit. Namely, two similarly extended portions of the
filament have been identified and show the following observed
emission-line ratios: [\ion{N}{ii}]$\lambda$ 6583/H$\alpha$ = 0.97
$\pm$ 0.17 and 0.47 $\pm$ 0.10, [\ion{S}{ii}]$\lambda$
6716+6731/H$\alpha$ = 1.76 $\pm$ 0.29 and 1.33 $\pm$ 0.19,
[\ion{S}{ii}] $\lambda$ 6716/[\ion{S}{ii}]$\lambda$ 6731 = 1.46
$\pm$ 0.26 and 1.64 $\pm$ 0.29 for the eastern and western
portion, respectively. The [\ion{O}{i}]$\lambda$ 6300 line
is visible in the bidimensional spectrum, after sky-subtraction,
at the same spatial position along the slit as the other emission lines,
but it is very faint. For this reason,
we measured it summing the spectrum over the whole extent 
of the filament, and found [\ion{O}{i}]/H$\alpha$ = 0.23 $\pm$ 0.41
and [\ion{N}{ii}]$\lambda$ 6583/[\ion{O}{i}] = 2.98 $\pm$ 0.46. 
The spectra of the two parts of the filament are plotted on
top of each other in Fig.~\ref{spectra} for easier comparability.
The splitting of the filament in two parts is clearly visible also
in Figs. 1 and 2, where there actually appear to be two crossing
filaments. The high value of the [\ion{S}{ii}] ratio, close to
1.5, prevents the determination of the electronic density N$_{\rm
e}$ and indicates that the excited gas is in the low density limit
N$_{\rm e}$ $<$100 cm$^{-3}$. The high ratios of low ionization
emission lines ([\ion{N}{ii}] and [\ion{S}{ii}]) to H$\alpha$ are
typical of shock heated gas \citep[e.g. ][]{bl97}. 
A high [\ion{N}{ii}]/H$\alpha$ ratio can also be produced by an
overabundance of nitrogen, which cannot be excluded in this case.
The emission lines are narrow and
unresolved (i.e. their width is instrumental, FWHM $\approx$ 5
\AA), thus indicating a low velocity shock.

We may roughly estimate the shock velocity. The non-detection of
the [\ion{O}{iii}] $\lambda$ 5007 line enables us to give a first approximate
upper limit for the shock velocity of 60 km s$^{-1}$ \citep{ray,fkd}. 
Then, by using the diagrams presented in \citet{hmr} for
a preshock density of 100 cm$^{-3}$ and a small magnetic field (0 - 1 $\mu$G) 
we estimated the shock velocity from the
[\ion{S}{ii}]$\lambda$6716+6731/H$\alpha$ ratio of $>$1.3.
According to these diagrams, our observed [\ion{S}{ii}]/H$\alpha$ ratio
implies an ionization fraction $\lesssim$ 0.03 and a shock velocity in the
range $\sim$  20 - 35 km s$^{-1}$. Also, the high [\ion{S}{ii}] $\lambda$
 6716/[\ion{S}{ii}]$\lambda$ 6731 ratio
indicative of the low density limit approaches the expectations for
slow shocks. Interestingly, the diagrams
in Fig.~8 of \citeauthor{hmr} show that, at such low shock velocities, 
the H$\alpha$/H$\beta$ ratio rises rapidly. This effect results from
the relative excitation rates at temperatures that are relatively low, 
2$\times$10$^4$ K or less. In particular, 
H$\alpha$/H$\beta$ $\gtrsim$ 5, the estimated lower limit in our spectra, 
corresponds to shock velocities $\lesssim$ 35 km s$^{-1}$, in agreement
with what derived from the [\ion{S}{ii}]/H$\alpha$ ratio.
Therefore, the slow shock, rather than a local dust patch, could be
the explanation to the high value of the Balmer decrement and
solve the discrepancy with the low foreground extinction value
derived from \citeauthor{sch}'s maps.

However, we note that our observed spectrum does not fully agree
with the models presented by \citet{hmr}. In particular, while
slow shock models predict high values ($\gtrsim$ 3) of
[\ion{N}{i}]$\lambda$ 5198+5201/[\ion{N}{ii}]$\lambda$ 6583
and low values ($\lesssim$ 0.1) of [\ion{N}{ii}]$\lambda$ 6583/[\ion{O}{i}]$\lambda$ 6300,
in our spectra the [\ion{N}{ii}] line is rather bright, while the [\ion{O}{i}] line
is very weak and the [\ion{N}{i}] lines are not detected. 
In fact the observed [\ion{O}{i}]/H$\alpha$ ratio would imply an ionization fraction
one order of magnitude higher than estimated from [\ion{S}{ii}]/H$\alpha$
and a shock velocity of $\approx$ 50 km s$^{-1}$, and the [\ion{N}{ii}]/[\ion{O}{i}]
ratio would imply an even higher shock velocity ($\sim$  80 km s$^{-1}$).
A possibility is that we are observing a superposition of different ionization 
layers. Another possibility is that the shock is taking place in partially
ionized gas. Indeed, models of slow shocks in neutral gas, such as those of \citet{hmr}, 
predict the high H$\alpha$/H$\beta$ ratios, but also very weak  [\ion{N}{ii}]
emission. A shock of the same speed in ionized gas would enhance the 
[\ion{N}{ii}] emission significantly, but the H$\alpha$/H$\beta$ ratio would be 
the radiative recombination value. A shock in partially ionized gas, 
could probably reproduce both the observed  H$\alpha$/H$\beta$ and 
[\ion{N}{ii}]/H$\alpha$ ratios; however, to our knowledge, there are no published 
tabulations of models of such shocks.

%

For completeness, we also used our higher resolution spectrum to determine
the radial velocity of the filament.
To improve the precision of the measurement we adopted an improved
version of the wavelength calibration procedure briefly described
in \citet{st04}, which derives the dispersion curve by matching 
all detected night-sky emission lines with a template generated by
folding the Osterbrock Sky Spectrum\footnote{Preparation of
the Osterbrock Sky Spectrum files was supported by grant ATM-9714636 from the 
NSF CEDAR program; see http://www.nvao.org/NVAO/download/Osterbrock.html.} 
to the observed resolution.  
Separate measurements of radial velocities were
initially obtained for the eastern and western part of the filament
in order to check for possible velocity differences between the them.
[\ion{N}{ii}]$\lambda$ 6583, H$\alpha$, [\ion{N}{ii}]$\lambda$ 6548,
[\ion{S}{ii}]$\lambda$6716 and 6731 lines were used. Resulting radial
velocities were V$_{\rm LSR}$ = $-$39 $\pm$ 4 km s$^{-1}$ and $-$33 $\pm$ 9 km s$^{-1}$
for the eastern and western part, respectively. 
Since no significant difference in velocity was detected,
we proceeded to an optimal extraction of the spectrum for
the whole extent of the nebular emission. This yielded 
a radial velocity V$_{\rm LSR}$ = $-$36.9 $\pm$ 0.8 km s$^{-1}$.
The reported error is the formal one from the line fits.
An inspection of the position of sky lines in the 
wavelength calibrated spectrum suggests that a more realistic estimate
of the error is a few km s$^{-1}$.
 
%

\subsection{A filament of the Monogem Ring?}

There are three main arguments why we believe that our optical
filament is a part of the MR. First, its location at the
south-eastern borders of the MR's X-ray image; second, because the
filament represents  a nebula excited by a slow shock, as expected
for this very extended SNR, which is close to pressure equilibrium
with the ISM; third, since we could not find an alternative origin
for the filament.

\begin{figure*}
   \centering
   \includegraphics{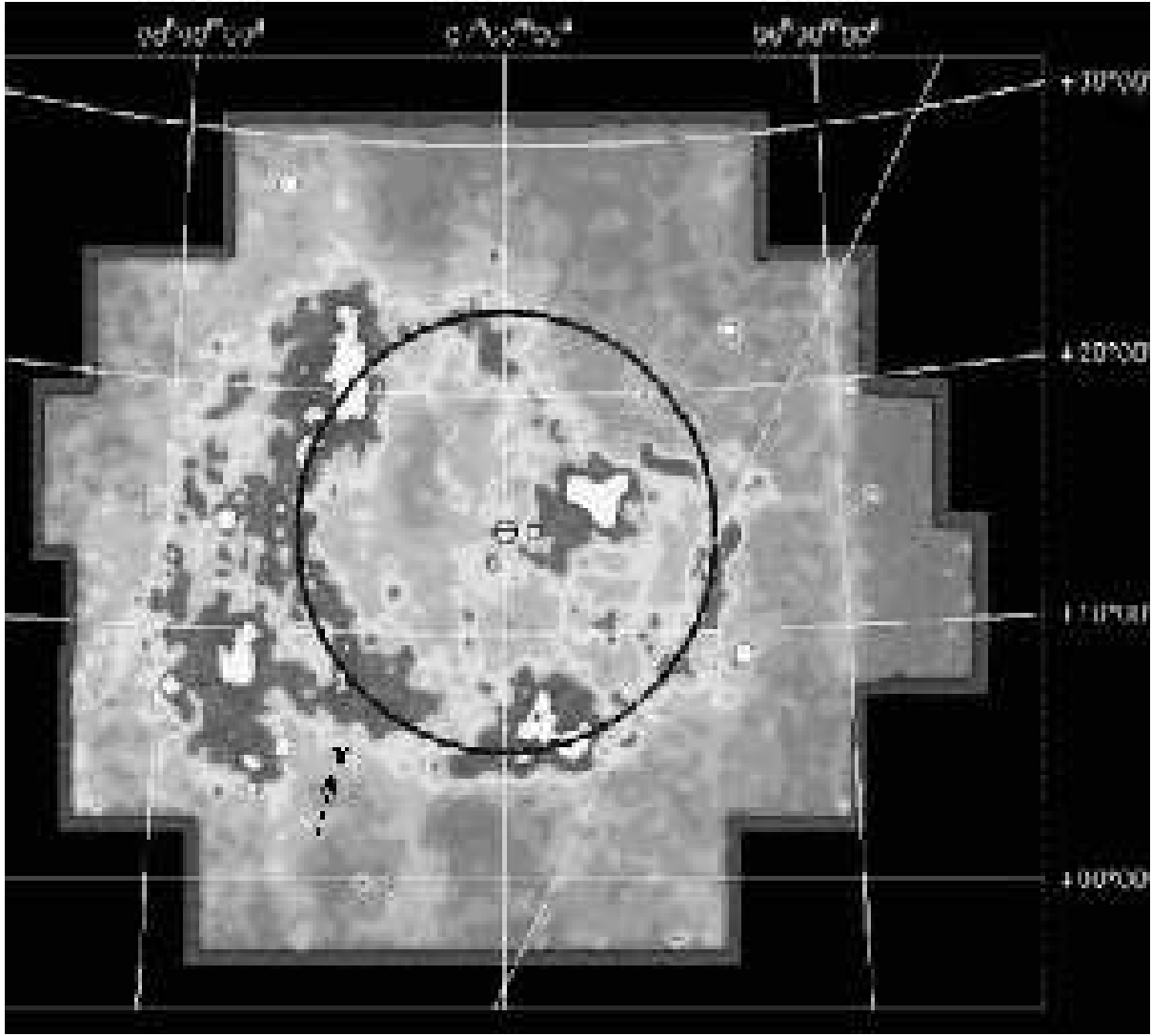}  
   \caption{  The Monogem Ring supernova remnant, as seen in the ROSAT
all-sky survey in the 0.25--0.75 keV X-ray band. The pulsar PSR
B0656+14 is marked with cross-hairs, and the $\sim$ 9${^\circ}$
circle shows the primary ring structure. The image is taken from
\citep{tbb}. The cross in the south-east, highlighted
by an arrow, shows the position of the new, $\sim$ 20\arcmin \,long
faint optical filament detected by us.}
              \label{Monogem_Ring}
    \end{figure*}

The first argument is obvious by examining
Fig.~\ref{Monogem_Ring}, where the location of the filament is
shown by a cross. In \citet{psa} a detailed comparison
of the MR with  e.g. $N_{\rm H}$ and IRAS maps is presented.
There, the region at and around the position of the filament is
entirely inconspicuous, i.e. it neither coincides with  an
enhancement or obvious lack of $N_{\rm H}$ or dust.

The second argument speaking in favour of a filament-MR
association has been described above. We note that the line
intensities and line ratios are strikingly similar to an other
faint nebula which obviously is the outcome of the interaction of
the very near (130 pc) and highly evolved SNR ``Orion-Eridanus
Bubble`` with a small ISM cloudlet \citep{zw}; the
main difference is, that the latter nebula with its intricate
morphology is projected towards the centre of the SNR.

 The third argument rests on
our check of pulsars within a 20\degr\, radius around the
filament's position by use of the ATNF Pulsar 
Catalogue\footnote{http://www.atnf.csiro.au/research/pulsar/psrcat}
\citep{mhth}: none of the one dozen pulsars (except PSR
B0656+14, the pulsar associated with the MR) comes into question,
either because of age and /or proper motion vector and/or
distance. Also, there is no sign of any star forming region within
many degrees. Further, using SIMBAD, we found no star in this
region which might have blown a wind and might have created a
bow-shock.

However, two objections against an association of the optical
filament with the MR can be produced. First, the above mentioned
lower limit to the foreground extinction of E($B-V$) $\sim$ 0.5
mag. This value is uncomfortably high and  - assuming that the
value is reliable - rather speaks in favour of a much larger
distance (than the distance of 300 pc for the MR) or in favour of
the presence of a cold small dust cloud in the foreground (of say,
15-20 K) in order to be absent or very faint even at 100 $\mu$m.
Dust formed in situ, coincident with or very close to this
filament cannot be excluded too. If the dust is in the foreground
and has an angular extension of several arcmin, it must however be
in the distance range 150 pc to 300 pc, since the F4V star GAT
1096 mentioned above appears to be unreddened  and is about 150 pc
distant. Nevertheless, as noticed in Sect. 3.2, the slow
shock could be responsible for an increase of the Balmer decrement
up to (or even above) our estimated value, without the need of
assuming the presence of dust. A few discrepancies between 
some observed emission-line ratios and the expectations from
slow shock models remain to be explained, though.

The second, less severe, objection is the overall orientation of
the filament (N-S) and its curvature; from both one would not be
able to deduce, even approximately, the centre of the MR. However,
even a quick examination of SNRs teeming with optical filaments
(like the Cygnus Loop) shows that, at the borders, one finds
filaments showing all kinds of alignments and curvature. In fact,
even modest variations in local density and shock velocity can
lead to such phenomena.

Hence, we consider it as
probable that we deal with a previously unexpected manifestation
of a part of the MR in the optical. The $\sim$ 20\arcmin \,long and
$\sim$ 1\arcmin \, wide filament has, at 300 pc distance, a length
of $\sim$ 1.7 pc and a width of $\sim$ 0.09 pc. However, a more
appropriate morphological description probably is, that we deal
with a thin sheet of emission seen edge-on. The sub-sheets,
particularly well visible in the southern straight part of the
filament, might be useful for future attempts to determine their
proper motion(s). A comparison of the 2m Schmidt telescope images
with POSS\,II (DSS2) did not lead to any observable proper motion
of the filament. Since the filament is located at the X-ray limb-brightened rim of
the MR, its velocity seems to be almost entirely in the plane of the sky.
For the epoch difference of 14 years between POSS\, II and the CCD imaging,
an assumed tangential velocity of 100 km s$^{-1}$, and a distance of 300 pc, 
a hard to detect proper motion of $\sim$  1\arcsec \, (i.e. about 1 pixel) would follow.
If the tangential velocity were much bigger, a
systematic shift in such an extended feature would be obvious.
Therefore, the lack of any noticeable proper motion agrees with
the slow shock interpretation.


%

%

\section{Conclusions}
We presented a previously unknown faint interstellar optical
filament, located at or projected onto the south-eastern borders
of the huge X-ray enhancement known as the Monogem Ring. This Ring
is a huge, nearby supernova remnant of high astrophysical
importance in several respects. Any optical emission was thought
to be practically undetectable or very faint, since this SNR will
hardly reach the radiative stage due to its expansion into ambient
matter of extraordinarily low density. The location of the optical
filament and the emission line ratios, which indicate the presence
of a low velocity shock, render this tenuous filament in all
probability to be an outlying portion of this evolved supernova
remnant. Using wide-angle CCD imaging could pay off in searches
for additional optical filaments along the borders of the Monogem
Ring.

\begin{acknowledgements}
      We are grateful to Dr. J. Raymond, for his useful 
      suggestions concerning shocks in partially ionized gas.
      This work was supported by the Austrian Science Fund (FWF),
      projects number P15316 and P17772. We acknowledge the use 
      of NASA's {\it SkyView} facility (http://skyview.gsfc.nasa.gov)
      located at NASA Goddard Space Flight Center and the use of the
      Southern H-Alpha Sky Survey Atlas (SHASSA), which is supported
      by the National Science Foundation. This research has made use
      of the SIMBAD database, operated at CDS, Strasbourg, France.
\end{acknowledgements}

\end{document}